\documentclass[12pt]{article}
\usepackage{amsfonts}
\usepackage{amsmath}
\usepackage{amsthm}
\usepackage{graphicx}
\usepackage{epstopdf}
\usepackage[lofdepth,lotdepth]{subfig}
\usepackage{caption}
\usepackage{float}
\usepackage[right=2.1cm,left=2.1cm, top=2.5cm, bottom=2.5cm]{geometry}

\title{MCMC for non-linear state space models \\ using ensembles of latent sequences}
\author{Alexander Y. Shestopaloff \\
Department of Statistical Sciences \\
University of Toronto \\
alexander@utstat.utoronto.ca \\
\and Radford M. Neal \\
Department of Statistical Sciences, \\
Department of Computer Science \\
University of Toronto \\
radford@utstat.utoronto.ca}
\date{30 April 2013}

\begin{document}

\maketitle

Non-linear state space models are a widely-used class of models for biological, economic, and physical processes. Fitting these models to observed data is a difficult inference problem that has no straightforward solution. We take a Bayesian approach to the inference of unknown parameters of a non-linear state model; this, in turn, requires the availability of efficient Markov Chain Monte Carlo (MCMC) sampling methods for the latent (hidden) variables and model parameters. Using the ensemble technique of Neal (2010) and the embedded HMM technique of Neal (2003), we introduce a new Markov Chain Monte Carlo method for non-linear state space models. The key idea is to perform parameter updates conditional on an enormously large ensemble of latent sequences, as opposed to a single sequence, as with existing methods. We look at the performance of this ensemble method when doing Bayesian inference in the Ricker model of population dynamics. We show that for this problem, the ensemble method is vastly more efficient than a simple Metropolis method, as well as $1.9$ to $12.0$ times more efficient than a single-sequence embedded HMM method, when all methods are tuned appropriately. We also introduce a way of speeding up the ensemble method by performing partial backward passes to discard poor proposals at low computational cost, resulting in a final efficiency gain of $3.4$ to $20.4$ times over the single-sequence method.

\section{Introduction}

Consider an observed sequence $z_{1}, \ldots, z_{N}$. In a state space model for $Z_{1}, \ldots, Z_{N}$, the distribution of the $Z_{i}$'s is defined using a latent (hidden) Markov process $X_{1}, \ldots, X_{N}$. We can describe such a model in terms of a distribution for the first hidden state, $p(x_{1})$, transition probabilities between hidden states, $p(x_{i} | x_{i-1})$, and emission probabilities, $p(z_{i} | x_{i})$, with these distributions dependent on some unknown parameters $\theta$.

While the state space model framework is very general, only two state space models, Hidden Markov Models (HMM's) and linear Gaussian models have efficient, exact inference algorithms. The forward-backward algorithm for HMM's and the Kalman filter for linear Gaussian models allow us to perform efficient inference of the latent process, which in turn allows us to perform efficient parameter inference, using an algorithm such as Expectation-Maximizaton for maximum likelihood inference, or various MCMC methods for Bayesian inference. No such exact and efficient algorithms exist for models with a continuous state space with non-linear state dynamics, non-Gaussian transition distributions, or non-Gaussian emission distributions, such as the Ricker model we consider later in this paper.

In cases where we can write down a likelihood function for the model parameters conditional on latent and observed variables, it is possible to perform Bayesian inference for the parameters and the latent variables by making use of sampling methods such as MCMC. For example, one can perform inference for the latent variables and the parameters by alternately updating them according to their joint posterior.

Sampling of the latent state sequence $x = (x_{1}, \ldots, x_{N})$ is difficult for state space models when the state space process has strong dependencies --- for example, when the transitions between states are nearly deterministic. To see why, suppose we sample from $\pi(x_{1}, \ldots, x_{N} | z_{1}, \ldots, z_{N})$ using Gibbs sampling, which samples the latent variables one at a time, conditional on values of other latent variables, the observed data, and the model parameters. In the presence of strong dependencies within the state sequence, the conditional distribution of a latent variable will be highly concentrated, and we will only be able to change it slightly at each variable update, even when the marginal distribution of this latent variable, given $z_{1}, \ldots, z_{N}$, is relatively diffuse. Consequently, exploration of the space of latent sequences will be slow.

The embedded HMM method of (Neal (2003), Neal, et al. (2004)) addresses this problem by updating the entire latent sequence at once. The idea is to temporarily reduce the state space of the model, which may be countably infinite or continuous, to a finite collection of randomly generated ``pool'' states at each time point. If the transitions between states are Markov, this reduced model is an HMM, for which we can use the forward-backward algorithm to efficiently sample a sequence with values in the pools at each time point. Pool states are chosen from a distribution that assigns positive probability to all possible state values, allowing us to explore the entire space of latent sequences in accordance with their exact distribution. Neal (2003) showed that when there are strong dependencies in the state sequence, the embedded HMM method performs better than conventional Metropolis methods at sampling latent state sequences.

In our paper, we first look at an MCMC method which combines embedded HMM updates of the hidden state sequence with random-walk Metropolis updates of the parameters. We call this method the `single-sequence' method. We next reformulate the embedded HMM method as an ensemble MCMC method. Ensemble MCMC allows multiple candidate points to be considered simultaneously when a proposal is made. This allows us to consider an extension of the embedded HMM method for inference of the model parameters when they are unknown. We refer to this extension as the ensemble embedded HMM method. We then introduce and describe a ``staged'' method, which makes ensemble MCMC more efficient by rejecting poor proposals at low computational cost after looking at a part of the observed sequence. We use the single-sequence, ensemble, and ensemble with staging methods to perform Bayesian inference in the Ricker model of population dynamics, comparing the performance of these new methods to each other, and to a simple Metropolis sampler.

\section{Ensemble MCMC}

We first describe Ensemble MCMC, introduced by Neal (2010) as a general MCMC method, before describing its application to inference in state space models.

Ensemble MCMC is based on the framework of MCMC using a temporary mapping. Suppose we want to sample from a distribution $\pi$ on $\mathcal{X}$. This can be done using a Markov chain with transition probablity from $x$ to $x'$ given by $T(x' | x)$, for which $\pi$ is an invariant distribution --- that is, $T$ must satisfy $\int \pi(x) T(x' | x) dx = \pi(x')$. The temporary mapping strategy defines $T$ as a composition of three stochastic mappings. The current state $x$ is stochastically mapped to a state $y \in \mathcal{Y}$ using the mapping $\hat{T}(y | x)$. Here, the space $\mathcal{Y}$ need not be the same as the space $\mathcal{X}$. The state $y$ is then updated to $y'$ using the mapping $\bar{T}(y' | y)$, and finally a state $x'$ is obtained using the mapping $\check{T}(x' | y')$. In this approach, we may choose whatever mappings we want, so long as the overall transition $T$ leaves $\pi$ invariant. In particular, if $\rho$ is a density for $y$, $T$ will leave $\pi$ invariant if the following conditions hold.
\begin{align}
&\int \pi(x) \hat{T}(y | x) dx = \rho(y) \\
&\int \rho(y) \bar{T}(y' | y) dy = \rho(y') \\
&\int \rho(y') \check{T}(x' | y') dy' = \pi(x')
\end{align}

In the ensemble method, we take $\mathcal{Y} = \mathcal{X}^{K}$, with $y = (x^{(1)}, \ldots, x^{(K)})$ referred to as an ``ensemble'', with $K$ being the number of ensemble elements. The three mappings are then constructed as follows. Consider an ``ensemble base measure'' over ensembles $(x^{(1)}, \ldots, x^{(K)})$ with density $\zeta(x^{(1)}, \ldots, x^{(K)})$, and with marginal densities $\zeta_{k}(x^{(k)})$ for each of the $k = 1, \ldots, K$ ensemble elements. We define $\hat{T}$ as
\begin{equation}
\hat{T}(x^{(1)}, \ldots, x^{(K)} | x) = \frac{1}{K} \sum_{k = 1}^{K} \zeta_{-k | k}(x^{(-k)} | x)\delta_{x}(x^{(k)})
\end{equation}
Here, $\delta_{x}$ is a distribution that places a point mass at $x$, $x^{(-k)}$ is all of $x^{(1)}, \ldots, x^{(K)}$ except $x^{(k)}$, and $\zeta_{-k | k}(x^{(-k)} | x^{(k)}) = \zeta(x^{(1)}, \ldots, x^{(K)}) / \zeta_{k}(x^{(k)})$ is the conditional density of all ensemble elements except the $k$-th, given the value $x^{(k)}$ for the $k$-th.

This mapping can be interpreted as follows. First, we select an integer $k$, from a uniform distribution on $\{1, \ldots, K\}$. Then, we set the ensemble element $x^{(k)}$ to $x$, the current state. Finally, we generate the remaining elements of the ensemble using the conditional density $\zeta_{-k | k}$.

The ensemble density $\rho$ is determined by $\pi$ and $\hat{T}$, and is given explicitly as
\begin{align}
\rho(x^{(1)}, \ldots, x^{(K)}) &= \int \pi(x) \hat{T}(x^{(1)}, \ldots, x^{(K)} | x) dx \notag \\
&=\zeta(x^{(1)}, \ldots, x^{(K)}) \frac{1}{K}\sum_{k = 1}^{K}\frac{\pi(x^{(k)})}{\zeta_{k}(x^{(k)})}
\end{align}

$\bar{T}$ can be any update (or sequence of updates) that leaves $\rho$ invariant. For example, $\bar{T}$ could be a Metropolis update for $y$, with a proposal drawn from some symmetrical proposal density. Finally, $\check{T}$ maps from $y'$ to $x'$ by randomly setting $x'$ to $x^{(k)}$ with $k$ chosen from $\{1, \ldots, K\}$ with probabilities proportional to $\pi(x^{(k)}) / \zeta_{k}(x^{(k)})$.

The mappings descibed above satisfy the necessary properties to make them a valid update, in the sense of preserving the stationary distribution $\pi$. The proof can be found in Neal (2010).

\section{Embedded HMM MCMC as an \\ Ensemble MCMC method}

The embedded HMM method briefly described in the introduction was not initially introduced as an ensemble MCMC method, but it can be reformulated as one. We assume here that we are interested in sampling from the posterior distribution of the state sequences, $\pi(x_{1}, \ldots, x_{N} | z_{1}, \ldots, z_{N})$, when the parameters of the model are known. Suppose the current state sequence is $x = (x_{1}, \ldots, x_{N})$. We want to update this state sequence in a way that leaves $\pi$ invariant.

The first step of the embedded HMM method is to temporarily reduce the state space to a finite number of possible states at each time, turning our model into an HMM. This is done by, for each time $i$, generating a set of $L$ ``pool'' states, $P_{i} = \{x_{i}^{[1]}, \ldots, x_{i}^{[L]}\}$, as follows. We first set the pool state $x_{i}^{[1]}$ to $x_{i}$, the value of the current state sequence at time $i$. The remaining $L - 1$ pool states $x_{i}^{[l]}$, for $l > 1$ are generated by sampling independently from some pool distribution with density $\kappa_{i}$. The collections of pool states at different times are selected independently of each other. The total number of sequences we can then construct using these pool states, by choosing one state from the pool at each time, is $K = L^{N}$.

The second step of the embedded HMM method is to choose a state sequence composed of pool states, with the probability of such a state sequence, $x$, being proportional to
\begin{equation}
q(x | z_{1}, \ldots, z_{N}) \textnormal{ } \propto \textnormal{ } p(x_{1})\prod_{i = 2}^{N}p(x_{i} | x_{i-1}) \prod_{i = 1}^{N}\biggl[\frac{p(z_{i} | x_{i})}{\kappa_{i}(x_{i})}\biggr]
\end{equation}
We can define $\gamma(z_{i} | x_{i}) = p(z_{i} | x_{i})/\kappa_{i}(x_{i})$, and rewrite $(6)$ as
\begin{equation}
q(x | z_{1}, \ldots, z_{N}) \textnormal{ } \propto \textnormal{ } p(x_{1})\prod_{i = 2}^{N}p(x_{i} | x_{i-1}) \prod_{i = 1}^{N}\gamma(z_{i} | x_{i})
\end{equation}
We now note that the distribution $(7)$ takes the same form as the distribution over hidden state sequences for an HMM in which each $x_{i} \in P_{i}$ --- the initial state distribution is proportional to $p(x_{1})$, the transition probabilities are proportional to $p(x_{i} | x_{i-1})$, and the $\gamma(z_{i} | x_{i})$ have the same role as emission probabilities. This allows us to use the well-known forward-backward algorithms for HMM's (reviewed by Scott (2002)) to efficiently sample hidden state sequences composed of pool states. To sample a sequence with the embedded HMM method, we first compute the ``forward'' probabilities. Then, using a stochastic ``backwards'' pass, we select a state sequence composed of pool states. (We can alternately compute backward probabilities and then do a stochastic forward pass). We emphasize that having an efficient algorithm to sample state sequences is crucial for the embedded HMM method. The number of possible sequences we can compose from the pool states, $L^{N}$, can be very large, and so naive sampling methods would be impractical.

In detail, for $x_{i} \in P_{i}$, the forward probabilities $\alpha_{i}(x_{i})$ are computed using a recursion that goes forward in time, starting from $i = 1$. We start by computing $\alpha_{1}(x_{1}) = p(x_{1})\gamma(z_{1} | x_{1})$. Then, for $1 < i \leq N$, the forward probabilities $\alpha_{i}(x)$ are given by the recursion
\begin{equation}
\alpha_{i}(x_{i}) = \gamma(z_{i} | x_{i})\sum_{l = 1}^{L} p(x_{i}| x^{[l]}_{i-1})\alpha_{i-1}(x^{[l]}_{i-1}), \quad \textnormal{for } x \in P_{i}
\end{equation}
The stochastic backwards recursion samples a state sequence, one state at a time, beginning with the state at time $N$. First, we sample $x_{N}$ from the pool $P_{N}$ with probabilities proportional to $\alpha_{N}(x_{N})$. Then, going backwards in time for $i$ from $N - 1$ to $1$, we sample $x_{i}$ from the pool $P_{i}$ with probabilities proportional to $p(x_{i+1} | x_{i})\alpha_{i}(x_{i})$, where $x_{i+1}$ is the variable just sampled at time $i + 1$. Both of these recursions are commonly implemented using logarithms of probabilities to avoid numerical underflow.

Let us now reformulate the embedded HMM method as an ensemble MCMC method. The step of choosing the pool states can be thought of as performing a mapping $\hat{T}$ which takes a single hidden state sequence $x$ and maps it to an ensemble of $K$ state sequences $y = (x^{(1)}, \ldots, x^{(K)})$, with $x = x^{(k)}$ for some $k$ chosen uniformly from $\{1, \ldots, K\}$. (However, we note that in this context, the order in the ensemble does not matter.) \\

Since the randomly chosen pool states are independent under $\kappa_{i}$ at each time, and across time as well, the density of an ensemble of hidden state sequences in the ensemble base measure, $\zeta$, is defined through a product of $\kappa_{i}(x_{i}^{[l]})$'s over the pool states and over time, and is non-zero for ensembles consisting of all sequences composed from the chosen set of pool states. The corresponding marginal density of a hidden state sequence $x^{(k)}$ in the ensemble base measure is
\begin{equation}
\zeta_{k}(x^{(k)}) = \prod_{i = 1}^{N} \kappa_{i}(x_{i}^{(k)})
\end{equation}
Together, $\zeta$ and $\zeta_{k}$ define the conditional distribution $\zeta_{-k | k}$, which is used to define $\hat{T}$. The mapping $\bar{T}$ is taken to be a null mapping that keeps the ensemble fixed at its current value, and the mapping $\check{T}$ to a single state sequence is performed by selecting a sequence $x^{(k)}$ from the ensemble with probabilities given by $(7)$, in the same way as in the embedded HMM method.

\section{The single-sequence embedded HMM MCMC method}

Let us assume that the parameters $\theta$ of our model are unknown, and that we want to sample from the joint posterior distribution of state sequences $x = (x_{1}, \ldots, x_{N})$ and parameters $\theta = (\theta_{1}, \ldots, \theta_{P})$, with density $\pi(x, \theta | z_{1}, \ldots, z_{N})$. One way of doing this is by alternating embedded HMM updates of the state sequence with Metropolis updates of the parameters. Doing updates in this manner makes use of an ensemble to sample state sequences more efficiently in the presence of strong dependencies. However, this method only takes into account a single hidden state sequence when updating the parameters.

The update for the sequence is identical to that in the embedded HMM method, with initial, transition and emission densities dependent on the current value of $\theta$. In our case, we only consider simple random-walk Metropolis updates for $\theta$, updating all of the variables simultaneously.

Evaluating the likelihood conditional on $x$ and $z$, as needed for Metropolis parameter updates, is computationally inexpensive relative to updates of the state sequence, which take time proportional to $L^{2}$. It may be beneficial to perform several Metropolis parameter updates for every update of the state sequence, since this will not greatly increase the overall computational cost, and allows us to obtain samples with lower autocorrelation time.

\section{An ensemble extension of the embedded HMM method}

When performing parameter updates, we can look at all possible state sequences composed of pool states by using an ensemble $((x^{(1)}, \theta), \ldots, (x^{(K)}, \theta))$ that includes a parameter value $\theta$, the same for each element of the ensemble. The update $\bar{T}$ could change both $\theta$ and $x^{(k)}$, but in the method we will consider here, we only change $\theta$.

To see why updating $\theta$ with an ensemble of sequences might be more efficient than updating $\theta$ given a single sequence, consider a Metropolis proposal in ensemble space, which proposes to change $\theta$ for all of the ensemble elements, from $\theta$ to $\theta^{*}$. Such an update can be accepted whenever there are \textit{some} elements $(x^{(k)}, \theta^{*})$ in the proposed ensemble that make the ensemble density $\rho((x^{(1)}, \theta^{*}), \ldots, (x^{(K)}, \theta^{*}))$ relatively large. That is, it is possible to accept a move in ensemble space with a high probability as long as \textit{some} elements of the proposed ensemble, with the new $\theta^{*}$, lie in a region of high posterior density. This is at least as likely to happen as having a proposed $\theta^{*}$ together with a single state sequence $x$ lie in a region of high posterior density.

Using ensembles makes sense when the ensemble density $\rho$ can be computed in an efficient manner, in less than $K$ times as long as it takes to compute $p(x, \theta | z_{1}, \ldots, z_{N})$ for a single hidden state sequence $x$. Otherwise, one could make $K$ independent proposals to change $x$, which would have approximately the same computational cost as a single ensemble update, and likely be a more efficient way to sample $x$. In the application here, $K = L^{N}$ is enormous for typical values of $N$ when $L \geq 2$, while computation of $\rho$ takes time proportional only to $NL^{2}$.

In detail, to compute the ensemble density, we need to sum $q(x^{(k)}, \theta | z_{1}, \ldots, z_{N})$ over all ensemble elements $(x^{(k)}, \theta)$, that is, over all hidden sequences which are composed of the pool states at each time. The forward algorithm described above makes it possible to compute the ensemble density efficiently by summing the probabilities at the end of the forward recursion $\alpha_{N}(x_{N})$ over all $x_{N} \in P_{N}$. That is
\begin{equation}
\rho((x^{(1)}, \theta), \ldots, (x^{(K)}, \theta)) \textnormal{ } \propto \textnormal{ } \pi(\theta)\sum_{k=1}^{K} q(x^{(k)}, \theta | z_{1}, \ldots, z_{N}) = \pi(\theta)\sum_{l = 1}^{L}\alpha_{N}(x_{N}^{[l]})
\end{equation}
where $\pi(\theta)$ is the prior density of $\theta$.

The ensemble extension of the embedded HMM method can be thought of as using an approximation to the marginal posterior of $\theta$ when updating $\theta$, since summing the posterior density over a large collection of hidden sequences approximates integrating over all such sequences. Larger updates of $\theta$ may be possible with an ensemble because the marginal posterior of $\theta$, given the data, is more diffuse than the conditional distribution of $\theta$ given a fixed state sequence and the data. Note that since the ensemble of sequences is periodically changed, when new pool states are chosen, the ensemble method is a proper MCMC method that converges to the exact joint posterior, even though the set of sequences using pool states is restricted at each MCMC iteration.

The ensemble method is more computationally expensive than the single-sequence method --- it requires two forward passes to evaluate the ensemble density for two values of $\theta$ and one backward pass to sample a hidden state sequence, whereas the single-sequence method requires only a single forward pass to compute the probabilities for every sequence in our ensemble, and a single backward pass to select a sequence.

Some of this additional computational cost can be offset by reusing the same pool states to do multiple updates of $\theta$. Once we have chosen a collection of pool states, and performed a forward pass to compute the ensemble density at the current value of $\theta$, we can remember it. Proposing an update of $\theta$ to $\theta^{*}$ requires us to compute the ensemble density at $\theta^{*}$ using a forward pass. Now, if this proposal is rejected, we can reuse the stored value of the ensemble density $\theta$ when we make another proposal using the same collection of pool states. If this proposal is accepted, we can remember the ensemble density at the accepted value. Keeping the pool fixed, and saving the current value of the ensemble density therefore allows us to perform $M$ ensemble updates with $M+1$ forward passes, as opposed to $2M$ if we used a new pool for each update.

With a large number of pool states, reusing the pool states for two or more updates has only a small impact on performance, since with any large collection of pool states we essentially integrate over the state space. However, pool states must still be updated occasionally, to ensure that the method samples from the exact joint posterior.

\section{Staged ensemble MCMC sampling}

Having to compute the ensemble density given the entire observed sequence for every proposal, even those that are obviously poor, is a source of inefficiency in the ensemble method. If poor proposals can be eliminated at a low computational cost, the ensemble method could be made more efficient. We could then afford to make our proposals larger, accepting occasional large proposals while rejecting others at little cost.

We propose to do this by performing ``staged'' updates. First, we choose a part of the observed sequence that we believe is representative of the whole sequence. Then, we propose to update $\theta$ to a $\theta^{*}$ found using an ensemble update that only uses the part of the sequence we have chosen. If the proposal found by this ``first stage'' update is accepted, we perform a ``second stage'' ensemble update given the entire sequence, with $\theta^{*}$ as the proposal. If the proposal at the first stage is rejected, we do not perform a second stage update, and add the current value of $\theta$ to our sequence of sampled values. This can be viewed as a second stage update where the proposal is the current state --- to do such an update, no computations need be performed.

Suppose that $\rho_{1}$ is the ensemble density given the chosen part of the observed sequence, and $q(\theta^{*}|\theta)$ is the proposal density for constructing the first stage update. Then the acceptance probability for the first stage update is given by
\begin{equation}
\min\biggl(1, \frac{\rho_{1}((x^{(1)}, \theta^{*}), \ldots, (x^{(K)}, \theta^{*}))q(\theta|\theta^{*})}{\rho_{1}((x^{(1)}, \theta), \ldots, (x^{(K)}, \theta))q(\theta^{*}|\theta)}\biggr)
\end{equation}
If $\rho$ is the ensemble density given the entire sequence, the acceptance probability for the second stage update is given by
\begin{equation}
\min\Biggl(1, \frac{\rho((x^{(1)}, \theta^{*}), \ldots, (x^{(K)}, \theta^{*}))\min\biggl(1, \frac{\rho_{1}((x^{(1)}, \theta^{*}), \ldots, (x^{(K)}, \theta^{*}))q(\theta|\theta^{*})}{\rho_{1}((x^{(1)}, \theta), \ldots, (x^{(K)}, \theta))q(\theta^{*}|\theta)}\biggr)}{\rho((x^{(1)}, \theta), \ldots, (x^{(K)}, \theta))\min\biggl(1, \frac{\rho_{1}((x^{(1)}, \theta), \ldots, (x^{(K)}, \theta))q(\theta^{*}|\theta)}{\rho_{1}((x^{(1)}, \theta^{*}), \ldots, (x^{(K)}, \theta^{*}))q(\theta|\theta^{*})}\biggr)}\Biggr)
\end{equation}
Regardless of whether $\rho_{1}((x^{(1)}, \theta^{*}), \ldots, (x^{(K)}, \theta^{*}))q(\theta|\theta^{*}) < \rho_{1}((x^{(1)}, \theta), \ldots, (x^{(K)}, \theta))q(\theta^{*}|\theta)$ or vice versa, the above ratio simplifies to
\begin{equation}
\min\Biggl(1, \frac{\rho((x^{(1)}, \theta^{*}), \ldots, (x^{(K)}, \theta^{*}))\rho_{1}((x^{(1)}, \theta^{*}), \ldots, (x^{(K)}, \theta^{*}))q(\theta|\theta^{*})}{\rho((x^{(1)}, \theta), \ldots, (x^{(K)}, \theta))\rho_{1}((x^{(1)}, \theta), \ldots, (x^{(K)}, \theta))q(\theta^{*}|\theta)}\Biggr)
\end{equation}
Choosing a part of the observed sequence for the first stage update can be aided by looking at the acceptance rates at the first and second stages. We need the moves accepted at the first stage to also be accepted at a sufficiently high rate at the second stage, but we want the acceptance rate at the first stage to be low so the method will have an advantage over the ensemble method in terms of computational efficiency. We can also look at the `false negatives' for diagnostic purposes, that is, how many proposals rejected at the first step would have been accepted had we looked at the entire sequence when deciding whether to accept.

We are free to select any portion of the observed sequence to use for the first stage. We will look here at using a partial sequence for the first stage consisting of observations starting at $n_{1}$ and going until the end, at time $N$. This is appropriate for our example later in the paper, where we only have observations past a certain point in time.

For this scenario, to perform a first stage update, we need to perform a backward pass. As we perform a backward pass to do the first stage proposal, we save the vector of ``backward'' probabilities. Then, if the first stage update is accepted, we start the recursion for the full sequence using these saved backward probabilities, and compute the ensemble density given the entire sequence, avoiding recomputation of backward probabilities for the portion of the sequence used for the first stage.

To compute the backward probabilities $\beta_{i}(x_{i})$, we perform a backward recursion, starting at time $N$. We first set $\beta_{N}(x_{N})$ to $1$ for all $x_{N} \in P_{N}$. We then compute, for $n_{1} \leq i < N$
\begin{equation}
\beta_{i}(x_{i}) = \sum_{l = 1}^{L}p(x_{i + 1}^{[l]}|x_{i})\beta_{i+1}(x_{i+1}^{[l]})\gamma(z_{i+1}|x_{i+1}^{[l]})
\end{equation}
We compute the first stage ensemble density $\rho_{1}$ as follows
\begin{equation}
\rho_{1}((x^{(1)}, \theta), \ldots, (x^{(K)}, \theta)) = \pi(\theta)\sum_{l = 1}^{L}p(x_{n_{1}}^{[l]})p(y_{n_{1}} | x_{n_{1}}^{[l]})\beta_{n_{1}}(x_{n_{1}}^{[l]})
\end{equation}
We do not know $p(x_{n_{1}})$, but we can choose a substitute for it (which affects only performance, not correctness). One possibility is a uniform distrubution over the pool states at $n_{1}$, which is what we will use in our example below.

The ensemble density for the full sequence can be obtained by performing the backward recursion up to the beginning of the sequence, and then computing
\begin{equation}
\rho((x^{(1)}, \theta), \ldots, (x^{(K)}, \theta)) = \pi(\theta)\sum_{l = 1}^{L}p(x_{1}^{[l]})p(y_{1} | x_{1}^{[l]})\beta_{1}(x_{1}^{[l]})
\end{equation}

To see how much computation time is saved with staged updates, we measure computation time in terms of the time it takes to do a backward (or equivalently, forward) pass --- generally, the most computationally expensive operation in ensemble MCMC --- counting a backward pass to time $n_{1}$ as a partial backward pass.

Let us suppose that the acceptance rate for stage 1 updates is $a_{1}$. An ensemble update that uses the full sequence requires us to perform two backwards passes if we update the pool at every iteration, whereas a staged ensemble update will require us to perform
\begin{equation}
1 + \frac{N - n_{1}}{N - 1} + a_{1}\frac{n_{1} - 1}{N - 1}
\end{equation}
backward passes on average (counting a pass back only to $n_{1}$ as $(N - n_{1})/(N - 1)$ passes). The first term in the above formula represents the full backward pass for the initial value of $\theta$ that is always needed --- either for a second stage update, or for mapping to a new set of pool states. The second term represents the partial backward pass we need to perform to complete the first stage proposal. The third term accounts for having to compute the remainder of a backwards pass if we accept the first stage proposal --- hence it is weighted by the first stage acceptance rate.

We can again save computation time by updating the pool less frequently. Suppose we decide to update the pool every $M$ iterations. Without staging, the ensemble method would require a total of $M + 1$ forward (or equivalently, backward) passes. With staged updates, the expected number of backward passes we would need to perform is
\begin{equation}
1 + M\frac{N - n_{1}}{N - 1} + Ma_{1}\frac{n_{1} - 1}{N - 1}
\end{equation}
as can be seen by generalizing the expression above to $M > 1$.

\section{Constructing pools of states}

An important aspect of these embedded HMM methods is the selection of pool states at each time, which determines the mapping $\hat{T}$ from a single state sequence to an ensemble. In this paper, we will only consider sampling pool states independently from some distribution with density $\kappa_{i}$ (though dependent pool states are considered in (Neal, 2003).

One choice of $\kappa_{i}$ is the conditional density of $x_{i}$ given $z_{i}$, based on some pseudo-prior distribution $\eta$ for $x_{i}$ --- that is, $\kappa_{i}(x_{i}) \propto \eta(x_{i})p(z_{i} | x_{i})$. This distribution approximates, to some extent, the marginal posterior of $x_{i}$ given all observations. We hope that such a pool distribution will produce sequences in our ensemble which have a high posterior density, and as a result make sampling more efficient.

To motivate how we might go about choosing $\eta$, consider the following. Suppose we sample values of $\theta$ from the model prior, and then sample hidden state sequences, given the sampled values of $\theta$. We can then think of $\eta$ as a distribution that is consistent with this distribution of hidden state sequences, which is in turn consistent with the model prior. In this paper, we choose $\eta$ heuristically, setting it to what we believe to be reasonable, and not in violation of the model prior.

Note that the choice of $\eta$ affects only sampling efficiency, not correctness (as long as it is non-zero for all possible $x_{i}$). However, when we choose pool states in the ensemble method, we cannot use current values of $\theta$, since this would make an ensemble update non-reversible. This restriction does not apply to the single-sequence method since in this case $\bar{T}$ is null.

\section{Performance comparisons on a \\ population dynamics model}

We test the performance of the proposed methods on the Ricker population dynamics model, described by Wood (2003). This model assumes that a population of size $N_{i}$ (modeled as a real number) evolves as $N_{i+1} = r N_{i}\exp(-N_{i} + e_{i})$, with $e_{i}$ independent with a Normal$(0,\sigma^{2})$ distribution, with $N_{0} = 1$. We don't observe this process directly, but rather we observe $Y_{i}$'s whose distribution is Poisson$(\phi N_{i})$. The goal is to infer $\theta = (r, \sigma, \phi)$. This is considered to be a fairly complex inference scenario, as evidenced by the application of recently developed inference methods such as Approximate Bayesian Computation (ABC) to this model. (See Fearnhead, Prangle (2012) for more on the ABC approach.) This model can be viewed as a non-linear state space model, with $N_{i}$ as our state variable.

MCMC inference in this model can be inefficient for two reasons. First, when the value of $\sigma^{2}$ in the current MCMC iteration is small, consecutive $N_{i}$'s are highly dependent, so the distribution of each $N_{i}$, conditional on the remaining $N_{i}$'s and the data, is highly concentrated, making it hard to efficiently sample state sequences one state at a time. An MCMC method based on embedding an HMM into the state space, either the single-sequence method or the ensemble method, can potentially make state sequence sampling more efficient, by sampling whole sequences at once. The second reason is that the distribution of $\theta$, given a single sequence of $N_{i}$'s and the data, can be concentrated as well, so we may not be able to efficiently explore the posterior distribution of $\theta$ by alternately sampling $\theta$ and the $N_{i}$'s. By considering an ensemble of sequences, we may be able to propose and accept larger changes to $\theta$. This is because the posterior distribution of $\theta$ summed over an ensemble of state sequences is less concentrated than it is given a single sequence.

To test our MCMC methods, we consider a scenario similar to those considered by Wood (2010) and Fearnhead and Prangle (2012). The parameters of the Ricker model we use are $r = \exp(3.8), \sigma = 0.15, \phi = 2$. We generated $100$ points from the Ricker model, with $y_{i}$ only observed from time $51$ on, mimicking a situation where we do not observe a population right from its inception.

We put a Uniform$(0, 10)$ prior on $\log(r)$, a Uniform$(0, 100)$ prior on $\phi$, and a Uniform$[\log(0.1), 0]$ prior on $\log(\sigma)$. Instead of $N_{i}$, we use $M_{i} = \log(\phi N_{i})$ as our state variables, since we found that doing this makes MCMC sampling more efficient. With these state variables, our model can be written as
\begin{align}
M_{1} &\sim N(\log(r) + \log(\phi) - 1, \sigma^{2}) \\
M_{i} | M_{i-1} &\sim N(\log(r) + M_{i-1} - \exp(M_{i-1})/\phi, \sigma^{2}), \quad i = 2, \ldots, 100 \\
Y_{i} | M_{i} &\sim \textnormal{Poisson}(\exp(M_{i})), \quad i = 51, \ldots, 100
\end{align}
Furthermore, the MCMC state uses the logarithms of the parameters, $(\log(r), \log(\sigma), \log(\phi))$.

For parameter updates in the MCMC methods compared below, we used independent normal proposals for each parameter, centered at the current parameter values, proposing updates to all parameters at once. To choose appropriate proposal standard deviations, we did a number of runs of the single sequence and the ensemble methods, and used these trial runs to estimate the marginal standard deviations of the logarithm of each parameter. We got standard deviation estimates of $0.14$ for $\log(r)$, $0.36$ for $\log(\sigma)$, and $0.065$ for $\log(\phi)$. We then scaled each estimated marginal standard deviation by the same factor, and used the scaled estimates as the proposal standard deviations for the corresponding parameters. The maximum scaling we used was $2$, since beyond this our proposals would often lie outside the high probability region of the marginal posterior density.

We first tried a simple Metropolis sampler on this problem. This sampler updates the latent states one at a time, using Metropolis updates to sample from the full conditional density of each state $M_{t}$, given by
\begin{align}
p(m_{1} | y_{51}, \ldots, y_{100}, m_{-1}) &\propto p(m_{1})p(m_{2}|m_{1}) \\
p(m_{i} | y_{51}, \ldots, y_{100}, m_{-i}) &\propto p(m_{i}|m_{i-1})p(m_{i+1}|m_{i}), \quad 2 \leq i \leq 50 \\
p(m_{i} | y_{51}, \ldots, y_{100}, m_{-i}) &\propto p(m_{i}|m_{i-1})p(y_{i}|m_{i})p(m_{i+1}|m_{i}), \quad 51 \leq i \leq 99 \\
p(m_{100} | y_{51}, \ldots, y_{100}, m_{-100}) &\propto p(m_{100}|m_{99})p(y_{100}|m_{100})
\end{align}
We started the Metropolis sampler with parameters set to prior means, and the hidden states to randomly chosen values from the pool distributions we used for the embedded HMM methods below. After we perform a pass over the latent variables, and update each one in turn, we perform a Metropolis update of the parameters, using a scaling of $0.25$ for the proposal density.

The latent states are updated sequentially, starting from $1$ and going up to $100$. When updating each latent variable $M_{i}$, we use a Normal proposal distribution centered at the current value of the latent variable, with the following proposal standard deviations. When we do not observe $y_{i}$, or $y_{i} = 0$, we use the current value of $\sigma$ from the state times $0.5$. When we observe $y_{i} > 0$, we use a proposal standard deviation of $1/\sqrt{1/\sigma^{2} + y_{i}}$ (with $\sigma$ from the state). This choice can be motivated as follows. An estimate of precision for $M_{i}$ given $M_{i-1}$ is $1/\sigma^{2}$. Furthermore, $Y_{i}$ is Poisson$(\phi N_{i})$, so that Var$(\phi N_{i}) \approx y_{i}$ and Var$(\log(\phi N_{i})) = \textnormal{Var}(M_{i}) \approx 1/y_{i}$. So an estimate for the precision of $M_{i}$ given $y_{i}$ is $y_{i}$. We combine these estimates of precisions to get a proposal standard deviation for the latent variables in the case when $y_{i} > 0$.

We ran the Metropolis sampler for $6,000,000$ iterations from five different starting points. The acceptance rate for parameter updates was between about $10\%$ and $20\%$, depending on the initial starting value. The acceptance rate for latent variable updates was between $11\%$ and $84\%$, depending on the run and the particular latent variable being sampled.

\begin{figure}[p]
  \centering
    \includegraphics[width=\textwidth]{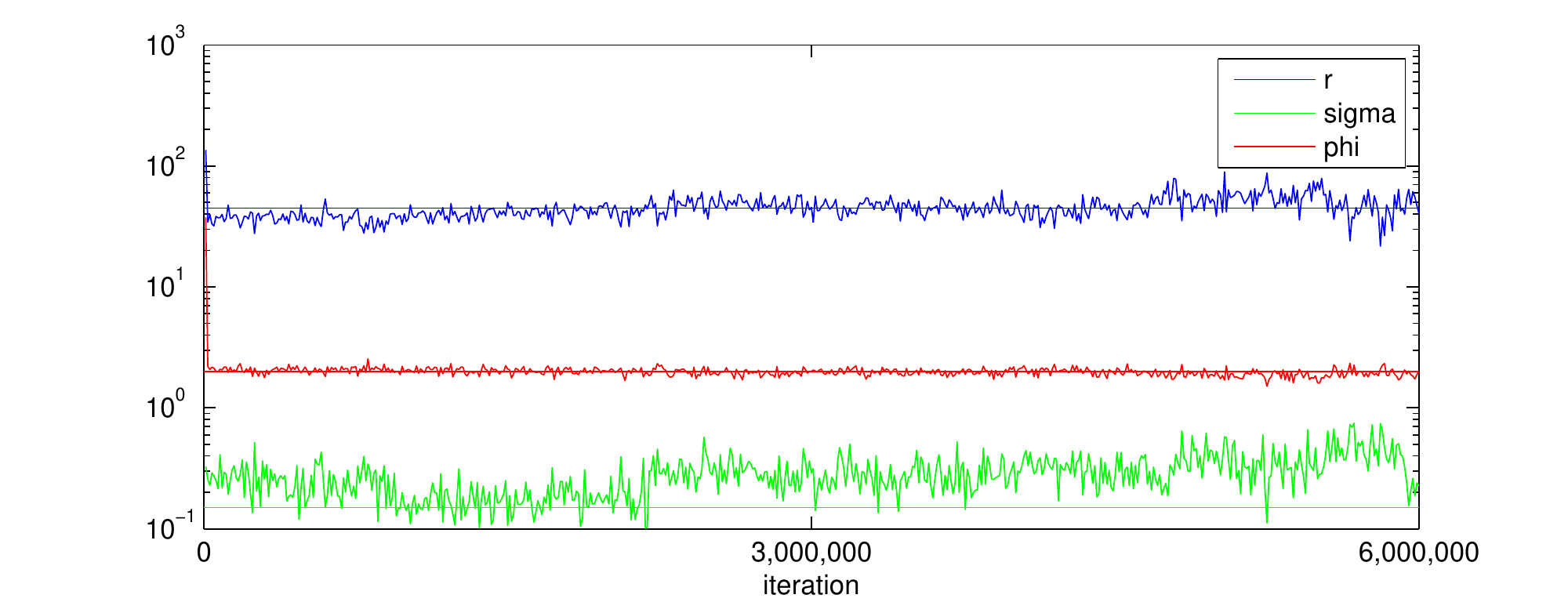}
    \caption{An example run of the simple Metropolis method.}\label{fig:ex-met}
\end{figure}

We found that the simple Metropolis sampler does not perform well on this problem. This is most evident for sampling $\sigma$. The Metropolis sampler appears to explore different regions of the posterior for $\sigma$ when it is started from different initial hidden state sequences. That is, the Metropolis sampler can get stuck in various regions of the posterior for extended periods of time. An example of the behaviour of the Metropolis sampler be seen on Figure \ref{fig:ex-met}. The autocorrelations for the parameters are so high that accurately estimating them would require a much longer run.

\begin{figure}[p]
  \centering
    \includegraphics[width=\textwidth]{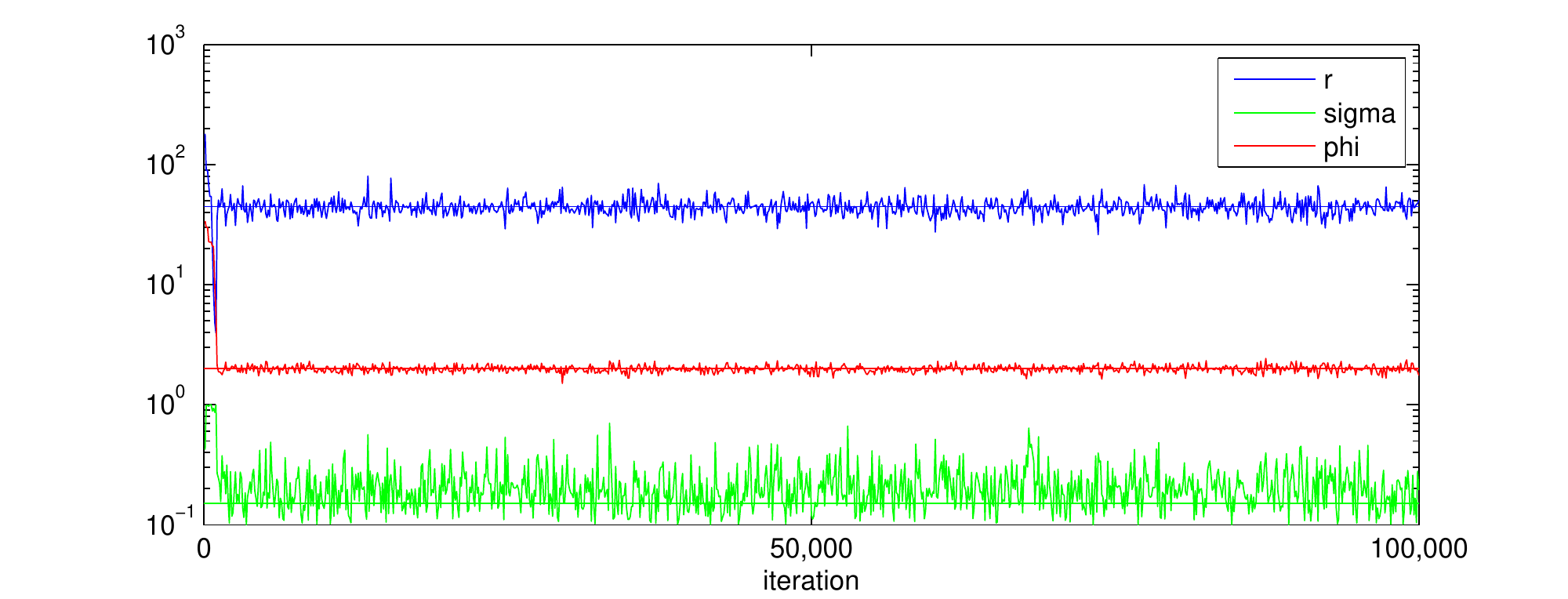}
    \caption{An example ensemble method run, with $120$ pool states and proposal scaling $1.4$.}\label{fig:ex-ens}
\end{figure}

This suggests that more sophisticated MCMC methods are necessary for this problem. We next looked at the single-sequence method, the ensemble method, and the staged ensemble method.

All embedded MCMC-based samplers require us to choose pool states for each time $i$. For the $i$'s where no $y_{i}$'s are observed, we choose our pool states by sampling values of $\exp(M_{i})$ from a pseudo-prior $\eta$ for the Poisson mean $\exp(M_{i})$ --- a Gamma$(k, \theta)$ distribution with $k = 0.15$ and $\theta = 50$ --- and then taking logarithms of the sampled values. For the $i$'s where we observe $y_{i}$'s we choose our pool states by sampling values of $\exp(M_{i})$ from the conditional density of $\exp(M_{i}) | y_{i}$ with $\eta$ as the pseudo-prior. Since $Y_{i} | M_{i}$ is Poisson$(\exp(M_{i}))$, the conditional density of $\exp(M_{i}) | y_{i}$ has a Gamma$(k + y_{i}, \theta/(1 + \theta))$ distribution.

It should be noted that since we choose our pool states by sampling $\exp(M_{i})$'s, but our model is written in terms of $M_{i}$, the pool density $\kappa_{i}$ must take this into account. In particular, when $y_{i}$ is unobserved, we have
\begin{equation}
\kappa_{i}(m_{i}) = \frac{1}{\Gamma(k)\theta^{k}}\exp(km_{i} - \exp(m_{i})/\theta), \quad -\infty < m_{i} < \infty
\end{equation}
and when $y_{i}$ is observed we replace $k$ with $k + y_{i}$ and $\theta$ with $\theta/(1 + \theta)$. We use the same way of choosing the pool states for all three methods we consider.

The staged ensemble MCMC method also requires us to choose a portion of the sequence to use for the first stage update. On the basis of several trial runs, we chose to use the last $20$ points, i.e. $n_{1} = 81$.

As we mentioned earlier, when likelihood evaluations are computationally inexpensive relative to sequence updates, we can do multiple parameter updates for each update of the hidden state sequence, without incurring a large increase in computation time. For the single-sequence method, we do ten Metropolis updates of the parameters for each update of the hidden state sequence. For the ensemble method, we do five updates of the parameters for each update of the ensemble. For staged ensemble MCMC, we do ten parameter updates for each update of the ensemble. These choices were based on numerous trial runs.

We thin each of the single-sequence runs by a factor of ten when computing autocorrelation times --- hence the time per iteration for the single-sequence method is the time it takes to do a single update of the hidden sequence and ten Metropolis updates. We do not thin the ensemble and staged ensemble runs.

To compare the performance of the embedded HMM methods, and to tune each method, we looked at the autocorrelation time, $\tau$, of the sequence of parameter samples, for each parameter. The autocorrelation time can be thought of as the number of samples we need to draw using our Markov chain to get the equivalent of one independent point (Neal (1993)). It is defined as
\begin{equation}
\tau = 1 + 2 \sum_{k=1}^{\infty}\rho_{k}
\end{equation}
where $\rho_{k}$ is the autocorrelation at lag $k$ for a function of state that is of interest. Here, $\hat{\rho}_{k} = \hat{\gamma}_{k} / \hat{\gamma}_{0}$, where $\hat{\gamma}_{k}$ is the estimated autocovariance at lag $k$. We estimate each $\hat{\gamma}_{k}$ by estimating autocovariances using each of the five runs, using the overall mean from the five runs, and then averaging the resulting autocovariance estimates. We then estimate $\tau$ by
\begin{equation}
\hat{\tau} = 1 + 2 \sum_{k=1}^{K}\hat{\rho}_{k}
\end{equation}
Here, the truncation point $K$ is where the remaining $\hat{\rho}_{k}$'s are nearly zero.

For our comparisons to have practical validity, we need to multiply each estimate of autocorrelation time estimate by the time it takes to perform a single iteration. A method that produces samples with a lower autocorrelation time is often more computationally expensive than a method that produces samples with a higher autocorrelation time, and if the difference in computation time is sufficiently large, the computationally cheaper method might be more efficient. Computation times per iteration were obtained with a program written in MATLAB on a Linux system with an Intel Xeon X5680 3.33 GHz CPU.

For each number of pool states considered, we started the samplers from five different initial states. Like with the Metropolis method, the parameters were initialized to prior means, and the hidden sequence was initialized to states randomly chosen from pool distribution at each time. When estimating autocorrelations, we discarded the first $10\%$ of samples of each run as burn-in.

An example ensemble run is shown in Figure \ref{fig:ex-ens}.  Comparing with Figure \ref{fig:ex-met}, one can see that the ensemble run has an enormously lower autocorrelation time.  Autocorrelation time estimates for the various ensemble methods, along with computation times, proposal scaling, and acceptance rates, are presented in Table \ref{table:results}. For the staged ensemble method, the acceptance rates are shown for the first stage and second stage. We also estimated the parameters of the model by averaging estimates of the posterior means from each of the five runs for the single-sequence method with $40$ pool states, the ensemble method with $120$ pool states, and the staged ensemble method with $120$ pool states. The results are presented in Table \ref{table:est}. The estimates using the three different methods do not differ significantly.

\begin{table}[t]
\small
\centering
\begin{tabular}{c|ccccc|ccc|ccc}
\hline
&         Pool  &         & Acc. & Iter- & Time /  & & ACT & & ACT & $\times$ & time\\ [0.5ex]
Method & states & Scaling & Rate &ations & iteration & $r$ & $\sigma$ & $\phi$ & $r$ & $\sigma$ & $\phi$ \\ [0.5ex]
\hline
		& 10 & & & 400,000 & 0.09 &            4345 &       2362 &               7272 & 391 &           213 & 654 \\ [1ex]
		& 20 & & & 400,000 & 0.17 &  \phantom{0}779 & 	    1875 &               1849 & 132 &           319 & 314 \\ [1ex]
Single-	 	& 40 & 0.25 & 12\% & 200,000 & 0.31 &  \phantom{0}427 & \phantom{0}187 & 1317 & 132 & \phantom{0}58 & 408 \\ [1ex]
sequence	& 60 & & & 200,000 & 0.47 &  \phantom{0}329 & \phantom{0}155 & \phantom{0}879 & 155 & \phantom{0}73 & 413 \\ [1ex]
		& 80 & & & 200,000 & 0.61 &  \phantom{0}294 & \phantom{0}134 & \phantom{0}869 & 179 & \phantom{0}82 & 530 \\
\hline
& 40 & 0.6 & 13\% & 100,000 & 0.23 &  \phantom{0}496 & \phantom{0}335 & \phantom{0}897 & 114 &  \phantom{0}77 & 206 \\ [1ex]
& 60 & 1 & 11\% & 100,000 & 0.34 &  \phantom{0}187 & \phantom{0}115 & \phantom{0}167 & \phantom{0}64 & \phantom{0}39 & \phantom{0}57 \\ [1ex]
Ensemble & 80 & 1 & 16\% & 100,000 & 0.47 &  \phantom{0}107 & \phantom{00}55 & \phantom{00}90 & \phantom{0}50 & \phantom{0}26 & \phantom{0}42 \\ [1ex]
& 120 & 1.4 & 14\% & 100,000 & 0.76 & \phantom{00}52 & \phantom{00}41 & \phantom{00}45 & \phantom{0}40 & \phantom{0}31 & \phantom{0}34 \\ [1ex]
& 180 & 1.8 & 12\% & 100,000 & 1.26 & \phantom{00}35 & \phantom{00}27 & \phantom{00}29 & \phantom{0}44 & \phantom{0}34 & \phantom{0}37 \\
\hline   
& 40 & 1 & 30, 15\% & 100,000 & 0.10 & \phantom{0}692 & \phantom{0}689 & 1201 & \phantom{0}69 & \phantom{0}69 & 120 \\ [1ex]
&  60 & 1.4 & 27, 18\% & 100,000 & 0.14 & \phantom{0}291 & \phantom{0}373 & \phantom{0}303 & \phantom{0}41 & \phantom{0}52 & \phantom{0}42 \\ [1ex]
Staged	 &  80 & 1.4 & 30, 25\% & 100,000 & 0.21 & \phantom{0}187 & \phantom{0}104 & \phantom{0}195 & \phantom{0}39 & \phantom{0}22 & \phantom{0}41 \\ [1ex]
Ensemble & 120 & 1.8 & 25, 29\% & 100,000 & 0.29 & \phantom{00}75 & \phantom{00}59 & \phantom{00}70 & \phantom{0}22 & \phantom{0}17 & \phantom{0}20 \\ [1ex]
& 180 & 2 & 23, 32\% & 100,000 & 0.45 & \phantom{00}48 & \phantom{00}44 & \phantom{00}52 & \phantom{0}22 & \phantom{0}20 & \phantom{0}23 \\
\hline
\end{tabular}
\caption{Comparison of autocorrelation times.}\label{table:results}
\end{table}

\begin{table}[b]
\small
\centering
\begin{tabular}{c c c c}
\hline
Method & $r$ & $\sigma$ & $\phi$ \\ [0.5ex]
\hline
Single-Sequence & 44.46 ($\pm$ 0.09) & 0.2074 ($\pm$ 0.0012) & 1.9921 ($\pm$ 0.0032) \\
Ensemble & 44.65 ($\pm$ 0.09) & 0.2089 ($\pm$ 0.0009) & 1.9853 ($\pm$ 0.0013) \\
Staged Ensemble & 44.57 ($\pm$ 0.04) & 0.2089 ($\pm$ 0.0010) & 1.9878 ($\pm$ 0.0015) \\
\hline
\end{tabular}
\caption{Estimates of posterior means, with standard errors of posterior means shown in brackets.}
\label{table:est}
\end{table}

From these results, one can see that for our Ricker model example, the single-sequence method is less efficient than the ensemble method, when both methods are well-tuned and computation time is taken into account. We also found that the the staged ensemble method allows us to further improve performance of the ensemble method.  In detail, depending on the parameter one looks at, the best tuned ensemble method without staging (with 120 pool states) is between 1.9 and 12.0 times better than the best tuned single-sequence method (with 40 pool states).  The best tuned ensemble method with staging (120 pool states) is between 3.4 and 20.4 times better than the best single-squence method.

The large drop in autocorrelation time for $\sigma$ for the single-sequence method between $20$ and $40$ pool states is due to poor mixing in one of the five runs. To confirm whether this systematic, we did five more runs of the single sequence method, from another set of starting values, and found that the same problem is again present in one of the five runs. This is inidicative of a risk of poor mixing when using the single-sequence sampler with a small number of pool states. We did not observe similar problems for larger numbers of pool states.

The results in Table \ref{table:results} show that performance improvement is greatest for the parameter $\phi$. One reason for this may be that the posterior distribution of $\phi$ given a sequence is significantly more concentrated than the marginal posterior distribution of $\phi$. Since for a sufficiently large number of pool states, the posterior distribution given an ensemble approximates the marginal posterior distribution, the posterior distribution of $\phi$ given an ensemble will become relatively more diffuse than the posterior distributions of $r$ and $\sigma$. This leads to a larger relative performance improvement when sampling values of $\phi$.

\begin{figure}[b!]
  \centering
    \includegraphics[width=\textwidth]{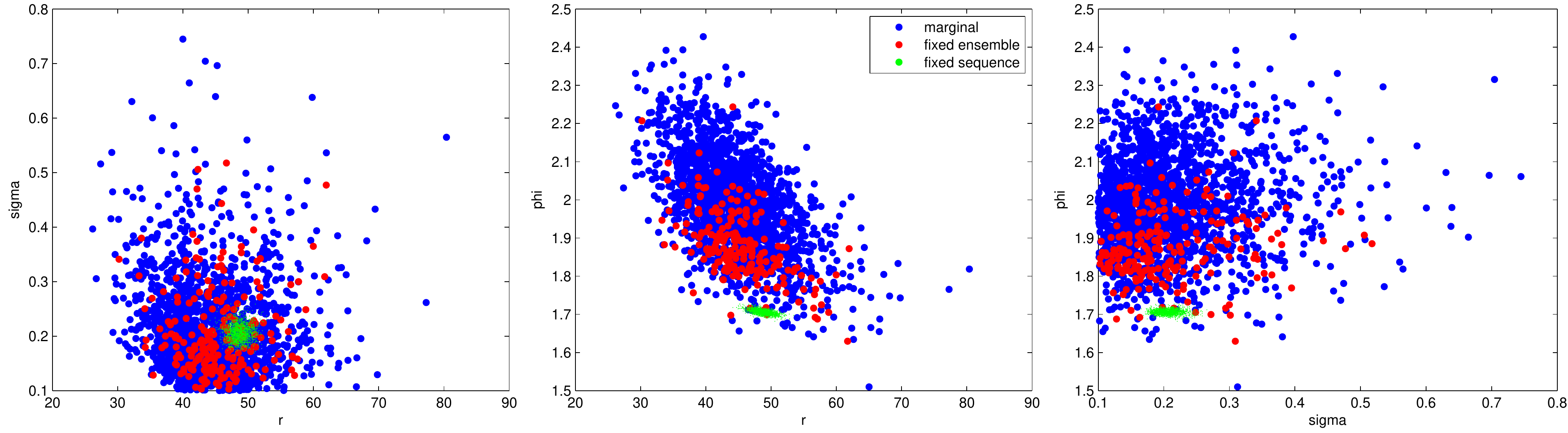}
    \caption{An illustration to aid explanation of relative performance. Note that the fixed sequence dots are smaller, to better illustrate the difference in the spread between the fixed sequence and two other distributions.}\label{fig:ill}
\end{figure}

Evidence of this can be seen on Figure \ref{fig:ill}. To produce it, we took the hidden state sequence and parameter values from the end of one of our ensemble runs, and performed Metropolis updates for the parameters, while keeping the hidden state sequence fixed. We also took the same hidden state sequence and parameter values, mapped the hidden sequence to an ensemble of sequences by generating a collection of pool states (we used $120$) and peformed Metropolis updates of the parameter part of the ensemble, keeping the pool states fixed. We drew $50,000$ samples given a single fixed sequence and $2,000$ samples given an ensemble.

Visually, we can see that the posterior of $\theta$ given a single sequence is significantly more concentrated than the marginal posterior, and the posterior given a fixed ensemble of sequences. Comparing the standard deviations of the posterior of $\theta$ given a fixed sequence, and the marginal posterior of $\theta$, we find that the marginal posterior of $\phi$ has a standard deviation about $21$ times larger than the posterior of $\phi$ given a single sequence. The marginal posteriors for $r$ and $\sigma$ have a posterior standard deviation larger by a factor of $5.2$ and $6.0$. The standard deviation of the posterior given our fixed ensemble of sequences is greater for $\phi$ by a factor of $11$, and by factors of $4.3$ and $3.2$ for $r$ and $\sigma$. This is consisent with our explanation above.

We note that the actual timings of the runs are different from what one may expect. In theory, the computation time should scale as $nL^{2}$, where $L$ is the number of pool states and $n$ is the number of observations. However, the implementation we use, in MATLAB, implements the forward pass as a nested loop over $n$ and over $L$, with another inner loop over $L$ vectorized. MATLAB is an interpreted language with slow loops, and vectorizing loops generally leads to vast performance improvements. For the numbers of pool states we use, the computational cost of the vector operation corresponding to the inner loop over $L$ is very low compared to that of the outer loop over $L$. As a result, the total computational cost scales approximately linearly with $L$, in the range of values of $L$ we considered. An implementation in a different language might lead to different optimal pool state settings.

The original examples of Wood and Fearnhead and Prangle used $\phi = 10$ instead of $\phi = 2$. We found that for $\phi = 10$, the ensemble method still performs better than the single sequence method, when computation time is taken into account, though the difference in performance is not as large. When $\phi$ is larger, the observations $y_{i}$ are larger on average as well. As a result, the data is more informative about the values of the hidden state variables, and the marginal posteriors of the model parameters are more concentrated. As a result of this, though we would still expect the ensemble method to improve performance, we would not expect the improvement to be as large as when $\phi = 2$.

Finally, we would like to note that if the single-sequence method is implemented already, implementing the ensemble method is very simple, since all that is needed is an additional call of the forward pass function and summing of the final forward probabilities. So, the performance gains from using an ensemble for parameter updates can be obtained with little additional programming effort.

\section{Conclusion}

We found that both the embedded HMM MCMC method and its ensemble extension perform significantly better than the ordinary Metropolis method for doing Bayesian inference in the Ricker model. This suggests that it would be promising to investigate other state space models with non-linear state dynamics, and see if it is possible to use the embedded HMM methods we described to perform inference in these models.

Our results also show that using staged proposals further improves the performance of the ensemble method. It would be worthwhile to look at other scenarios where this technique might be applied.

Most importantly, however, our results suggest that looking at multiple hidden state sequences at once can make parameter sampling in state space models noticeably more efficient, and so indicate a direction for further research in the development of MCMC methods for non-linear, non-Gaussian state space models.

\section*{Acknowledgements}

This research was supported by the Natural Sciences and Engineering Research Council of Canada.  A.~S.\ is in part funded by an NSERC Postgraduate Scholarship. R.~N.\ holds a Canada Research Chair in Statistics and Machine Learning.

\section*{References}

\leftmargini 0.2in
\labelsep 0in

\begin{description}

\item
Fearnhead, P., Prangle, D. (2012) ``Constructing summary statistics for approximate Bayesian computation: semi-automatic approximate Bayesian computation'', {\em Journal of the Royal Statistical Society, Series B}, vol.~74, pp.~1-28.

\item
Neal, R. M. (2010) ``MCMC Using Ensembles of States for Problems with Fast and Slow Variables such as Gaussian Process Regression'', Technical Report No. 1011, Department of Statistics, University of Toronto.

\item
Neal, R. M. (2003) ``Markov Chain Sampling for Non-linear State Space Models using Embedded Hidden Markov Models'', Technical Report No. 0304, Department of Statistics, University of Toronto.

\item
Neal, R. M., Beal, M. J., and Roweis, S. T. (2004) ``Inferring state sequences for non-linear systems with embedded hidden Markov models'', in S. Thrun, et al (editors), {\em Advances in Neural Information Processing Systems 16}, MIT Press.

\item
Neal, R. M. (1993) ``Probabilistic Inference Using Markov Chain Monte Carlo Methods'', Technical Report CRG-TR-93-1, Dept. of Computer Science, University of Toronto.

\item
Steven L. Scott (2002) ``Bayesian Methods for Hidden Markov Models: Recursive Computing in the 21st Century'', {\em Journal of the American Statistical Association}. vol.~97, no.~457, pp.~337-351.

\item
Wood, S. (2010) ``Statistical inference for noisy nonlinear ecological dynamic systems'', {\em Nature}. vol.~466, pp.~1102-1104.

\end{description}

\end{document}